\documentstyle[a4,umlaute,epsf,12pt]{article}

\newcounter{saveeqn}

\newcommand{\be}{\begin{equation}}
\newcommand{\ee}{\end{equation}}

\newcommand{\bea}{\begin{eqnarray}}
\newcommand{\eea}{\end{eqnarray}}
\begin{document}

\begin{center}

{\centerline{2nd March 2000} 
\centerline{Submitted to J. Rheol.}
{\bf Molecular Weight Dependent Kernels in Generalized Mixing Rules
}
}
{
\vskip 3.0 true mm
\noindent
\sc Wolfgang Thimm $^{1}$  \\
\sc Christian Friedrich $^{1}$ \footnote{Author to whom all correspondence
should be addressed. Electronic mail:chf@fmf.uni-freiburg.de}\\
\sc Tobias Roths $^{1}$ \\
\sc Josef Honerkamp $^{1,2}$ \\
\vskip 3.0 true mm
{\it $^{1}$ Freiburger Materialforschungszentrum, Stefan-Meier-Stra{\ss}e 21,
 D-79104 Freiburg im Breisgau, Germany}

\vskip 3.0 true mm
{\it $^{2}$ Universit\"at Freiburg, Fakult\"at f\"ur Physik, 
Hermann-Herder-Stra{\ss}e~3, D-79104 Freiburg im Breisgau, Germany}
}

\end{center}

\section*{Synopsis}

In this paper, a model is proposed for the kernel in the generalized 
mixing rule recently formulated by Anderssen and Mead (1998). In order 
to derive such a model, it is necessary to take account of the 
rheological significance of the kernel in terms of the relaxation 
behaviour of the individual polymers involved. This leads naturally 
to consider a way how additional physical effects, which depend on the
molecular weight distribution, can be included in 
the mixing rule.  
The advantage of this approach is that, without changing the generality 
derived by Anderssen and Mead (1998), the choice of the model 
proposed here for the kernel guarantees the enhanced physical 
and rheological significance of their mixing rule. 
 
\section*{I. Introduction}

In an earlier article (Thimm et al. (1999))
an analytical relationship was derived between the
relaxation time spectrum and the molecular weight distribution (MWD):
\be \label{rel} w(m) =  \frac{1}{\beta}\frac{{\alpha}^{(1/\beta)}}
{(G_N^{0})^{1/\beta}
} \tilde h(m)
 (\int_{ m}^{\infty} \frac{\tilde h(m')}{m'} 
{\rm d} m')^{(1/\beta-1)}.\ee
This relation is based on a 
generalized mixing rule formulated and analyzed by Anderssen and Mead (1998). 
Among other things it can be used to calculate the MWD $w(m)$ once an
estimate of the relaxation time spectrum $h(\tau)$ has been derived.
The corresponding inverse relationship is given by:
\be \label{inverse} 
\frac{\tilde h(m)}
{G_N^0}=\frac{\beta}{\alpha} w(m) [\int_m^\infty {\rm d} m' 
\frac{ w(m')}{m'}]^{\beta-1}
.\ee
In these equations, $\beta$ is the generalized mixing parameter,
$G_N^0$ the plateau shear modulus and $\alpha$ is the
scaling exponent 
\be \label{tau} \tau = k m^{\alpha}, k={\rm const.},\ee
where $\alpha \approx 3.4$,
which is normally determined experimentally.
Let
$\tilde h(m) \equiv h(\tau(m))$,
$m$ be the dimensionless molecular weight $m=M/M_0$, where  
$M_0$ is the monomer molecular 
weight and $M$ is the molecular weight of the polymer.
The above relationships are valid in the molecular 
weight range: $m_e < m < \infty$, where $m_e$ denotes the entanglement
molecular weight.

The generalized mixing rule, used in the derivation of the relationship Eq.(\ref{rel}),
was taken to have the form
\be \label{mi}
G(t)=G^0_N(\int^{\infty}_{m_e}F^{1/\beta}(t,m)\frac{w(m)}{m}
{\rm d} m)^{\beta}.
\ee
The value for $\beta$ 
can be determined theoretically from polymer dynamical considerations.
It takes the value 1
for single reptation (Doi and Edwards (1986)), and 2 for double
reptation or entanglement 
(Tsenoglou (1987,1991), Des Cloizeaux (1988)). Thimm et al. (2000) have shown
that a value of $\beta \approx 2$ can be justified, if the Rouse spectrum
is treated correctly in the evaluation of rheological data.
The relaxation processes
of the individual polymer chains (reptation, double reptation)
with molecular weight $m$, which correspond to the experimentally
determined relaxation shear modulus $G(t)$, are described
by the integral kernel $F(t,m)$. Based on either polymer
dynamical considerations or phenomenological observations,
five kernels have been examined and discussed in the literature
(see e.g. Wasserman and Graessly (1992), Maier et al. (1998)):
single exponential; Tuminello; Doi; Des Cloizeaux and
BSW (Baumg\"artel-Schausberger-Winter).
All of these kernels do not depend on the nature of the polymer being 
investigated. In Maier et al. (1998) the single
exponential kernel gave the best agreement with experimental observations.

Keeping the generality of the mixing rule Eq. (\ref{mi})
it is possible to derive
a kernel, which depends on $w(m)$ and has the single exponential kernel as the
limit for monodisperse distributions.
The derivation of this kernel is discussed in section II. The
physical relevance of the kernel is discussed in section III.

\section*{II. Derivation of the MWD-dependent kernel}

The decomposition of rheological material functions, such as the
relaxation shear modulus $G(t)$, in terms of Maxwell-modes is an accepted
procedure to present results of rheological measurements (see e.g. Ferry 
(1980)). The following
equation combines the relaxation shear modulus $G(t)$ with the corresponding
relaxation time spectrum $h(\tau)$:
\be \label{refh} G(t) = \int_{0}^{\infty} \frac{h(\tau)}
{\tau} e^{-t/\tau} {\rm d} \tau.\ee
The equilibrium shear modulus $G_e$ is assumed to be zero for
the viscoelastic liquids under consideration.
Whereas a decomposition in terms of
other kernels is possible too, the use of Eq. (\ref{refh})
has been established empirically. 

We summarize the
derivation of the results found in Thimm et al. (1999), where 
the definition Eq. (\ref{refh})
of the relaxation time spectrum was combined with the mixing
rule Eq. (\ref{mi}) above. 
The algebraic transformation:
\be \label{algeb}
\int_{a}^{\infty}{\rm d} x (-\frac{\rm d}{{\rm d} x})
[\int_x^{\infty}f(x')dx']^{\gamma}=[\int_a^{\infty}f(x)dx]^{\gamma}=\gamma \int_
a^{\infty}
f(x)[\int_x^{\infty}f(x')dx']^{\gamma-1} dx \ee
was applied on Eq. (\ref{refh}):
\begin{eqnarray}
G(t)^{1/\beta} = (\int_{\tau_e}^{\infty} \frac{h(\tau)}{\tau} 
e^{-t/\tau }d\tau)^{1/\beta}=
(\int_{m_e}^{\infty} \alpha
\frac{\tilde h(m')}{m'} e^{-t/\tau(m')} 
{\rm d}m')^{1/\beta}=  \nonumber \\
 \int_{m_e}^{\infty}\frac{\alpha^{(1/\beta)}}{\beta} \frac{\tilde h(m)}
 {m} e^{-t/\tau(m)
} 
[\int_{m}^{\infty} \frac{\tilde h(m')}{m'} e^{-t/\tau(m')} 
{\rm d}m']^{(1/\beta)-1} {\rm d} m  .
\end{eqnarray}
and the result inserted in Eq. (\ref{mi}). The following
relation connecting the relaxation time spectrum
and the MWD $w(m)$ was found:
\begin{eqnarray} \label{Zeit2}
 \int_{m_e}^{\infty}[\frac{a^{(1/\beta)}}{\beta} 
 \frac{\tilde h(m)}{m} e^{-t/\tau(m)} 
[\int_{m}^{\infty} \frac{\tilde h(m')}{m'} e^{-t/\tau(m')} 
{\rm d}m']^{(1/\beta)-1} \nonumber \\  -(G_N^0)^{1/\beta}F(t,m)^{1/\beta} 
\frac{w(m)}{m} 
] {\rm d} m 
 = 0.  
\end{eqnarray}

One solution, which fulfills Eq. (\ref{Zeit2}) is that the kernel under the
integral is identical to zero:
\begin{eqnarray} \label{Zeit4}
\frac{a^{(1/\beta)}}{\beta} 
 \tilde h(m) e^{-t/\tau(m)} 
[\int_{m}^{\infty} \frac{\tilde h(m')}{m'} e^{-t/\tau(m')} 
{\rm d}m']^{(1/\beta)-1} \nonumber \\  -(G_N^0)^{1/\beta}F(t,m)^{1/\beta} 
w(m) 
  = 0.  
\end{eqnarray}
For the time $t=0$, we have $F(t=0,m)=1$, because
$G(t=0)=G_N^0$ and the integral over the molecular weight distribution
in the mixing rule Eq. {\ref{mi}} becomes one.  

Hence we derive from Eq. (\ref{Zeit4}) for $t=0$ the equation (\ref{rel}). 
When the analytical relation (\ref{rel}) is inserted in
equation (\ref{Zeit4}), the constants (i.e. $\alpha, G_N^0$) cancel and one finds that
\be
\exp(-t/\tau)[\int_m^{\infty}\frac{\tilde h(m')}{m'} \exp(-t/\tau(m') 
{\rm d} m']^{1/\beta-1}= F(t,m)^{1/\beta}[\int_m^{\infty} \frac{\tilde h(m')}
{m'} {\rm d} m']^{1/\beta-1}.
\ee
This equation can be reordered (using Eq. (\ref{inverse}))
to find an integral kernel, which depends implicitly
on the molecular weight distribution: 
\be \label{iinver}
F(t,m)=\exp(-t/\tau(m))^{\beta}[\frac{\int_m^{\infty} 
\frac {w(m'')}{m''}(\int_{m''}^{\infty}\frac{w(m')}{m'}{\rm d} m')^{\beta-1} 
\exp(-t/\tau(m'')) {\rm d} m''}{\int^{\infty}_{m} \frac{w(m'')}{m''}
(\int_{m''}^{\infty}\frac{w(m')}{m'}{\rm d} m')^{\beta-1} {\rm d}m''}]
^{1-\beta}.
\ee
When $\beta=2$ (double reptation) is inserted
in Eq. (\ref{iinver}), one finds that
\be \label{breit}
F(t,m)=\exp(-t/\tau(m))\frac{\exp(-t/\tau(m)) \int_{m}^{\infty}
\frac{w(m'')}{m''}(\int_{m''}^{\infty}\frac{w(m')}{m'}{\rm d} m') {\rm d}m''}
{\int_m^{\infty} 
\frac {w(m'')}{m''}(\int_{m''}^{\infty}\frac{w(m')}{m'}{\rm d} m') 
\exp(-t/\tau(m'')){\rm d} m''}.
\ee
For monodisperse distributions the result that the kernel is of single
exponential type (single exponential kernel) 
is easily reconstructed (inserting Dirac's $\delta$-function:
$w(m)/m=\delta(m-m_0)$, where the
integration over the $\delta$-function gives 1).
This observation agrees with the result,
that the single exponential kernel describes the data
best, found by Maier et al. (1998)
evaluating rheological data of polystyrene mixtures. 
 For
other MWDs Eq.'s (\ref{iinver}, \ref{breit})
establish a new kind of integral kernel, to be used
in the mixing rule Eq. (\ref{mi}).

In the next section some features of the proposed kernel are discussed.

\section*{III. Discussion}

The physical interpretation of the kernel is that the relaxation of a single
polymer seems to be implicitly dependent on the molecular weight distribution
of the neighbouring polymers. While for a monodisperse MWD the
result of the single exponential kernel is found, the kernel predicts
a more complex behaviour for highly polydisperse polymer melts. 

There are several contributions in literature (e.g. Graessly (1982), Montfort
et al. (1984), Rubinstein and Colby (1987), McLeish(1992)),
which discuss that the relaxation behaviour is different in monodisperse
and polydisperse samples. The relaxation behaviour is accelerated in 
polydisperse samples. The reason is that when a short chain reptates
away, the entanglements between long chains become ineffective. The
short chains are considered as solvent, which dilute the density
of entanglements
in the polymer melt and therefore lead to an effective faster
relaxation process. The proposed kernel agrees qualitatively with this
picture.

The relaxation behaviour of the kernels is illustrated in Fig. 1.
We have used a monomodale log-normal distribution with 
common parameters and $\beta=2$ to simulate the
data in the shown figures. We find that the relaxation is faster in
the highly polydisperse than in the near monodisperse sample.

When the kernel is inserted in the mixing rule Eq.(\ref{mi}), it is found
that the main influence of the molecular weight distribution on
$G(t)$ is due to the integration over
$w(m)$, whereas the 
MWD in the kernels has a smaller influence on $G(t)$. In Fig. 2 are 
several $G(t)$ plotted, which were
simulated with either the pure exponential or with the proposed kernel.
The $G(t)$ curves are shifted mainly parallel to each other, but there are
no additional structural changes besides the shift.
One finds that the main influence of the MWD on $G(t)$ is independent of
the kernel's details. 
This observation is in agreement with the
idea introduced by Thimm et al. (1999),
that the molecular weight distribution can be
determined from rheological data without explicitly regarding the
details of the kernels.

To achieve a better understanding of Eq. (\ref{breit}), we
approximate the exponential in the integral by a step function
($\exp(-t/\tau(m)=\theta(\tau(m)-t)$), where 
$\theta(\tau(m)-t)=1$ for $t<\tau(m)$ and
$\theta(\tau(m)-t)=0$ for $t>\tau(m)$).
This approximation leads to a lower limit of the integral
in Eq. (\ref{breit}), which depends on the time and
is denoted by $\tilde m(t)$. With this simplification, one
can use Eq. (\ref{algeb}) and obtains a simpler form of the kernel:
\be \label{special}
F(t,m)= [\exp(-t/\tau) \frac{\int_m^{\infty} \frac{w(m')}{m'} {\rm d}
m'}{\int_{\tilde m(t)}^{\infty} \frac{w(m')}{m'} {\rm d}
m'}]^{2}.
\ee
Inserting this kernel in the mixing rule (\ref{mi}), one 
finds (for $\beta=2$) that:
\be \label{mineu}
G(t)=G^0_N{\Large[}\int^{\infty}_{m_e}\exp(-t/\tau(m))  [\frac{
\int_{m}^{\infty}\frac{w(m')}{m'} {\rm d}
m'}{\int_{\tilde m(t)}^{\infty} \frac{w(m')}{m'} {\rm d} m'} ]  \frac{w(m)}{m}
{\rm d} m{\Large ]}^{2}.
\ee
Besides the single exponential behaviour another term is found (in brackets),
which depends implicitly on the MWD and the time.

The mathematical details of the integration over
MWD as in Eq.
(\ref{special})
using typical MWDs as ansatz have been discussed by Eder et al. (1989). 
We do not think such details relevant for the study discussed in 
this article, but emphazise that the analytical relations (\ref{rel},
\ref{inverse}) are valid independent of any assumptions concerning
the MWD.

\section*{IV. Conclusion}

We have proposed a new kernel, which is implicitly dependent on the
molecular weight distribution.  In the limit of a monodisperse
sample this kernel contains the well-known result of the single exponential
kernel as special case. The physical
interpretation of this kernel is that the relaxation of a single
polymer seems to be implicitly dependent on the molecular weight distribution
of the neighbouring polymers.

This observation enlarges the physical features of the class of mixing
rules discussed by Anderssen and Mead (1998) in a new way.
The advantage of this implicitly molecular weight distribution
dependent kernel is that the 
physics (the reptation processes) described by the kernels 
can be fine tuned but the general form of the mixing rule
discussed by Anderssen and Mead (1998) stays the same.

\section*{Acknowledgement}

We thank M. Marth for helpful discussions and R. S. Anderssen for assistance in
formulation of the manuscript.
W. B. Thimm was supported by the Deutsche Forschungsgemeinschaft: Graduiertenkolleg f\"ur 
Strukturbildung in makromolekularen Systemen.


\section*{References}

\noindent
 Anderssen R. S. and D. W. Mead
``Theoretical derivation of molecular weight scaling for rheological
  parameters,''
 J.~Non-Newtonian Fluid Mech. {\bf 76}, 299-306 (1998).\\

\noindent
Des Cloizeaux, J.,
``Double reptation vs simple reptation in polymer melts,''
 Europhys. Lett. {\bf 5}, 437-442 (1988); {\bf 6}, 475 (1988).\\

\noindent
 Doi, M. and S. F. Edwards, {\it The Theory of Polymer Dynamics},
(Clarendon, Oxford 1986). \\

\noindent
Eder G., Janeshitz-Kriegl H., Liedauer S., Schausberger A.,
Stadlbauer W., ``The influence of Molar Mass Distribution
on the Complex Moduli of Polymer Melts'', J. Rheol. {\bf 33(6)}, 805-820
(1989) \\

\noindent
 Ferry, F. D.,  {\it Viscoelastic Properties of Polymers},
3rd ed. (Wiley, New York 1980).\\

\noindent
Graessly, W. W., ``Entangled linear, branched and Network Polymer systems
- molecular theories,'' Adv. Polym. Sci. 47 , 67-117 (1982) \\

\noindent
Maier, D., A. Eckstein, C. Friedrich, J. Honerkamp,
``Evaluation of Models Combining Rheological Data with the
Molecular Weight distribution,''
 J. Rheol. {\bf 42(5)}, 1153-1173 (1998) \\

\noindent
Montfort, J.-P., Marin, G., Monge, P., ``Effects of constraint release
on the dynamics of entangled linear polymer melts,''
Macromolecules {\bf 17}, 1551-1560 (1984) \\

\noindent
McLeish, T. C. B.,
``Relaxation behaviour of highly polydisperse polymer melts,''
Polymer {\bf 33 (13)}, 2852-2854 (1992) \\

\noindent
Rubinstein, M., Colby R. H., ``Self-consistent theory of polydisperse
entangled polymers: Linear viscoelasticity of binary blends,''
J. Chem. Phys. {\bf 89 (8)}, 5291-5306 (1988) \\

\noindent
Thimm, W., C. Friedrich, M. Marth, J. Honerkamp,
``An Analytical Relation between Relaxation Time Spectrum and
Molecular Weight Distribution, ''
J. Rheol. {\bf 43(6)}, 1663-1672 (1999)  \\

\noindent
Thimm, W., C. Friedrich, M. Marth, J. Honerkamp,
``On the Rouse spectrum and the determination of the molecular
weight distribution,''
J. Rheol. {\bf 44(2)}, 429-438 (2000)\\

\noindent
Tsenoglou, C.,
``Viscoelasticity of binary homopolymer blends,''
  ACS Polym. Prepr. {\bf 28}, 185-186 (1987).\\

\noindent
Tsenoglou, C., ``Molecular Weight Polydispersity Effects on the
Viscoelasticity of Entangled Linear Polymers,''
Macromolecules {\bf 24}, 1762-1767 (1991). \\

\noindent
 Wasserman, S. H. and W. W. Graessley,
``Effects of polydispersity on linear viscoelasticity
in entangled polymer melts,''
J. Rheol. {\bf 36}, 543-572 (1992).\\

\unitlength=1 cm

\newpage
{\large \bf FIG. 1}\\
For various polydisperse molecular weight distributions (Mw=300kg/mol,
Ip= Mw/Mn = 1.2,8)
the
behaviour of the proposed kernel at fixed $M \approx 1000 kg/mol$ is illustrated. Solid lines:
single 
exponential kernel, dashed line (Ip=1.2), dotted line (Ip=8).
\\
\begin{picture}(1,10.5)
  \epsfxsize=6.7cm
   \put(0,1.7){\epsffile{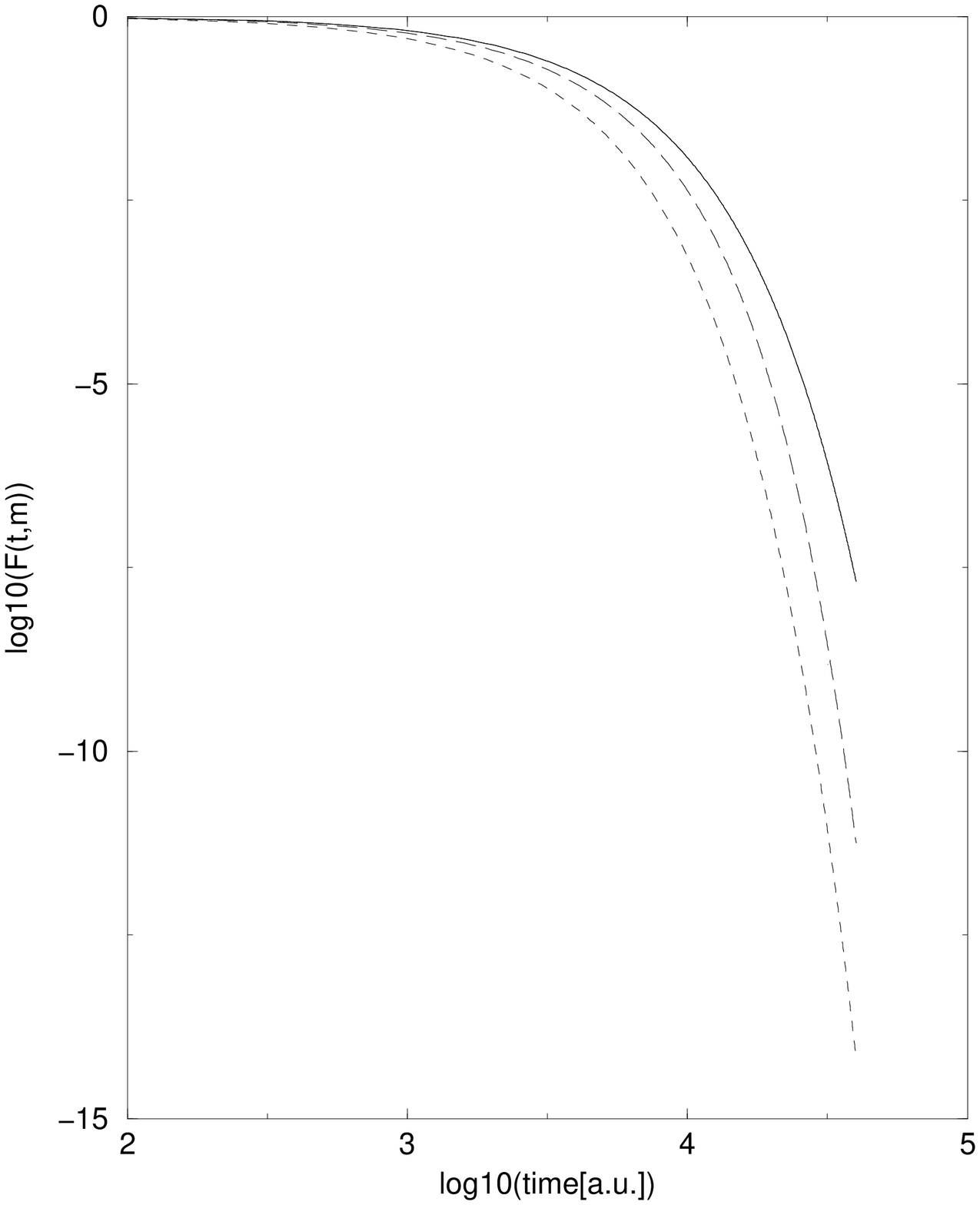}}
 \end{picture}

\unitlength=1 cm

\newpage
{\large \bf FIG. 2}\\
For two polydisperse molecular weight distributions 
Ip=1.2 and 8, Mw=300 kg/mol,
the
behaviour of $G(t)/G_N^0$ is illustrated. The two different kernels
are distinguished:
single exponential kernel (solid), the implicitly molecular weight
distribution dependent kernel (dashed).
\\
\begin{picture}(1,10.5)
  \epsfxsize=6.7cm
   \put(0,1.7){\epsffile{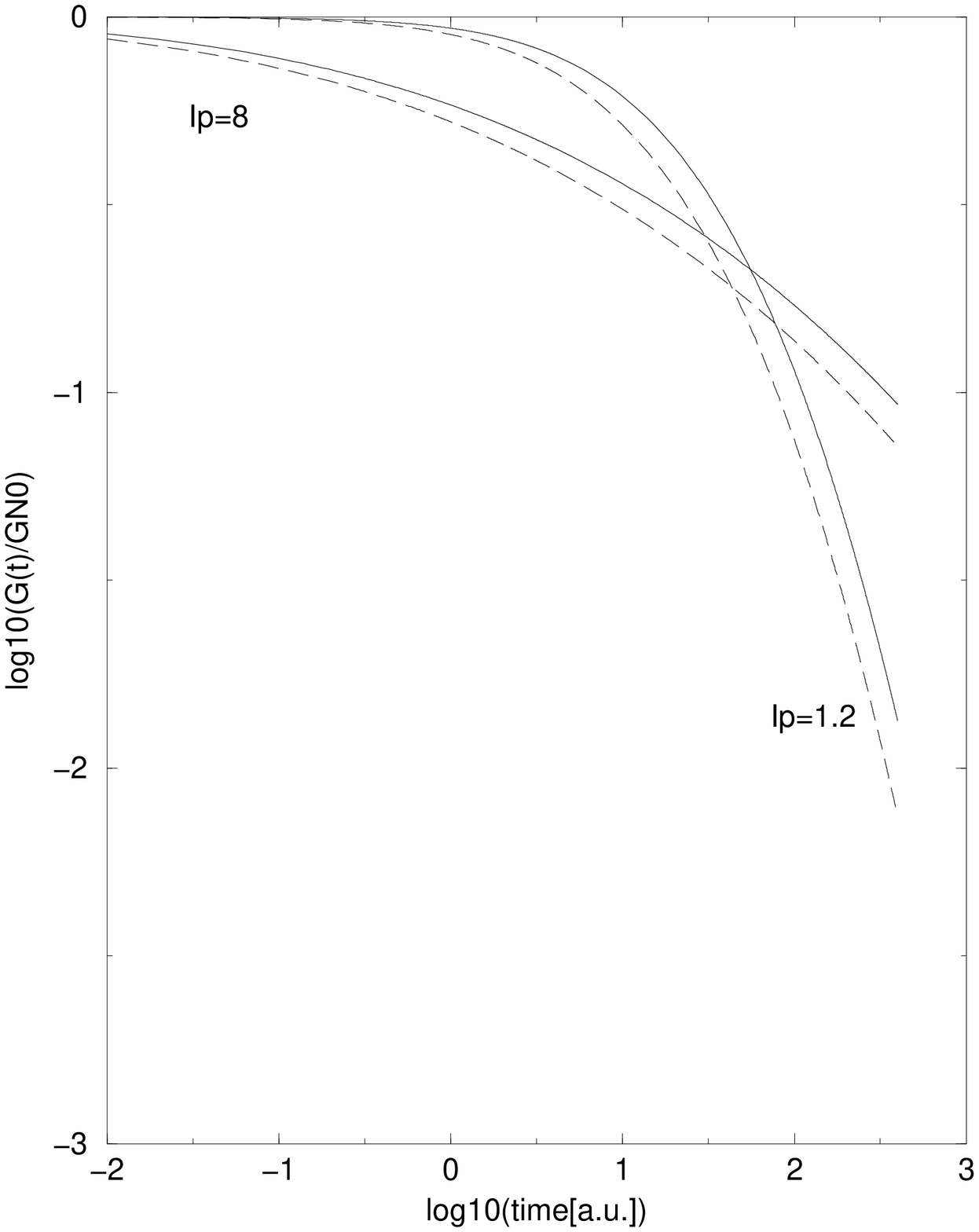}}
 \end{picture}

\end{document}